\documentclass[twocolumn,showpacs,showkeys,amsmath,amssymb,aps,pre]
{revtex4-1} 

\usepackage[final]{graphicx} 
\usepackage{epstopdf} \usepackage[dvipsnames]{xcolor} 
\usepackage{dcolumn} \usepackage{bm} 

\newcommand\etal{\mbox{\textit{et al.}}} \graphicspath{{./fig/}} 

\begin{document}

\title{Emergence of spatio-temporal dynamics from exact coherent 
solutions in pipe flow} 

\author{Paul Ritter$^{1,2}$, Fernando Mellibovsky$^{3}$ and Marc 
Avila$^{1,2}$} 

\affiliation{$^{1}$Institute of Fluid Mechanics, 
Friedrich-Alexander-Universit\"at Erlangen-N\"urnberg, 91058 
Erlangen, Germany \\$^2$Center of Applied Space Technology and 
Microgravity, University of Bremen, 28359 Bremen, Germany\\ 
$^{3}$Castelldefels School of Telecom and Aerospace Engineering, 
Universitat Polit\`ecnica de Catalunya, 08222 Barcelona, Spain} 

\date{\today} 

\begin{abstract} Turbulent-laminar patterns are ubiquitous near transition in 
wall-bounded shear flows. Despite recent progress in describing their 
dynamics in analogy to non-equilibrium phase transitions, there is no 
theory explaining their emergence. Dynamical-system approaches 
suggest that invariant solutions to the Navier--Stokes equations, 
such as traveling waves and relative periodic orbits in pipe flow, 
act as building blocks of the disordered dynamics. While recent 
studies have shown how transient chaos arises from such solutions, 
the ensuing dynamics lacks the strong fluctuations in size, shape and 
speed of the turbulent spots observed in experiments. We here show 
that chaotic spots with distinct dynamical and kinematic properties 
merge in phase space and give rise to the enhanced spatio-temporal 
patterns observed in pipe flow. This paves the way for a 
dynamical-system foundation to the phenomenology of turbulent-laminar 
patterns in wall-bounded extended shear flows.  \end{abstract} 

\maketitle 


\section{Introduction} 

Despite the ubiquity of turbulence, a theory explaining the emergence 
of its complexity has remained elusive even though the governing 
equations have been known for nearly two centuries. 
Landau~\cite{landau1944} proposed in the 1940s that as the Reynolds 
number Re increases, turbulence arises through an infinite sequence 
of linear instabilities. In this picture each instability would add a 
new temporal frequency to the flow eventually resulting in the 
continuous spectrum that is characteristic of turbulence. Ruelle and 
Takens~\cite{ruelle1971,*newhouse1978} refined this idea and 
established the current paradigm for the onset of turbulence: a few 
instabilities suffice to produce a continuous spectrum and render the 
flow chaotic. Such a scenario was first successfully demonstrated in 
laboratory experiments in a Taylor--Couette setup~\cite{gollub1975}, 
and illustrations of other routes to chaos via period-doubling and 
intermittency followed shortly 
after~\cite{feigenbaum1978,*pomeau1980,*eckmann1981,*libchaber1982}. 
This framework has been useful in explaining the origin of temporal 
chaos in many systems but falls short of describing turbulence 
because it fails to account for spatial features~\cite{chate1987}. 

\begin{figure} \centering 
\begin{tabular}{c} 
(a)\\ 
\includegraphics[width=0.75\linewidth]{fig_1a.eps}\\ 
(b)\\ 
\quad\includegraphics[width=0.71\linewidth]{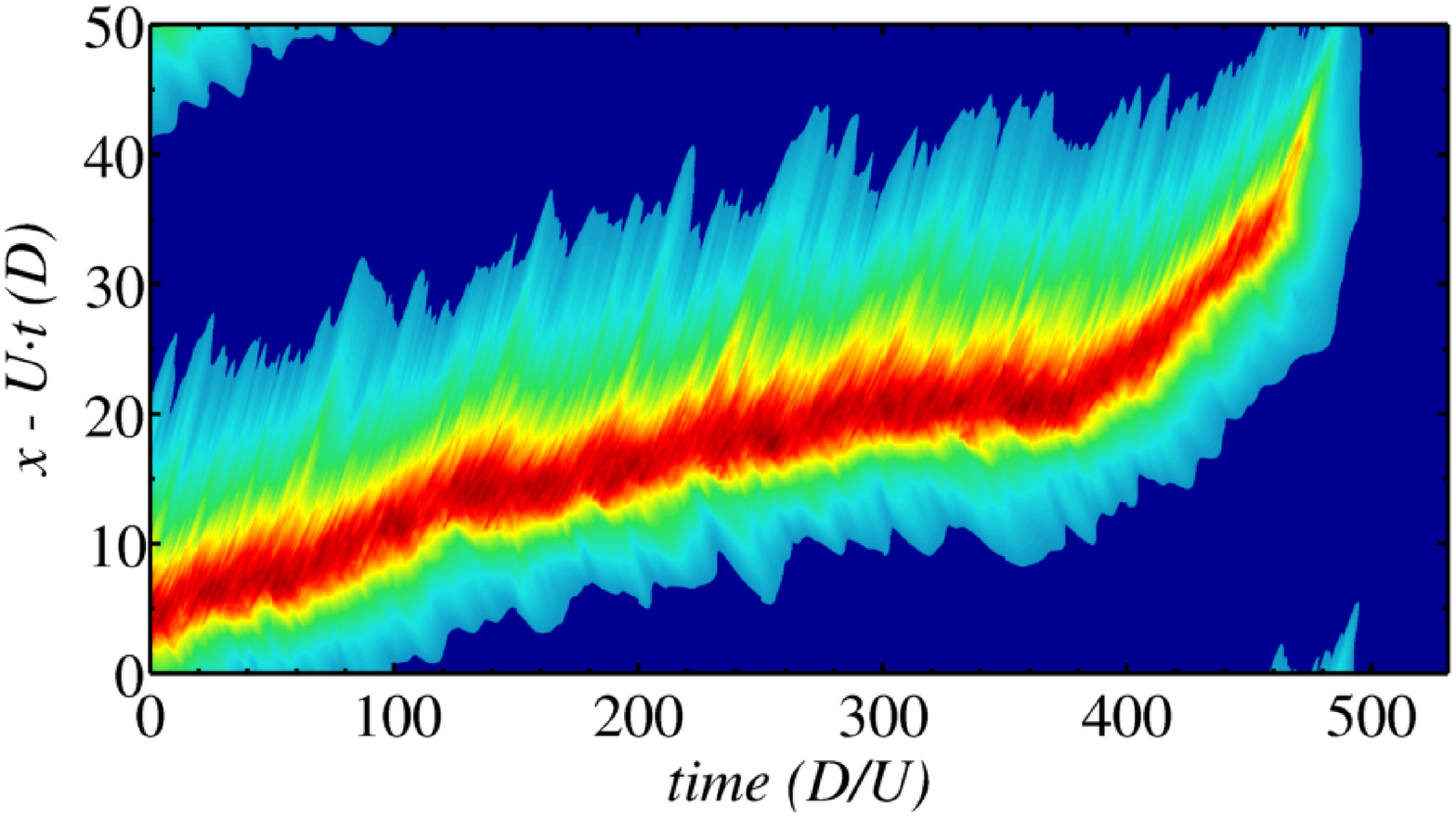}\\ 
(c)\\ 
\begin{tabular}{cc} 
$t=150\,D/U$ & $225\,D/U$ \\ 
\includegraphics[height=0.09\linewidth]{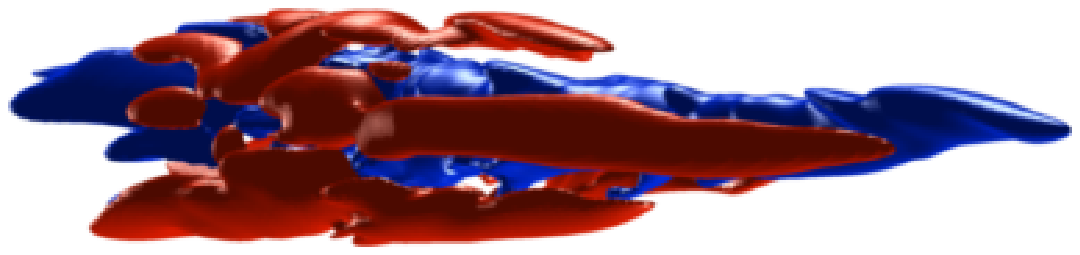}& 
\includegraphics[height=0.09\linewidth]{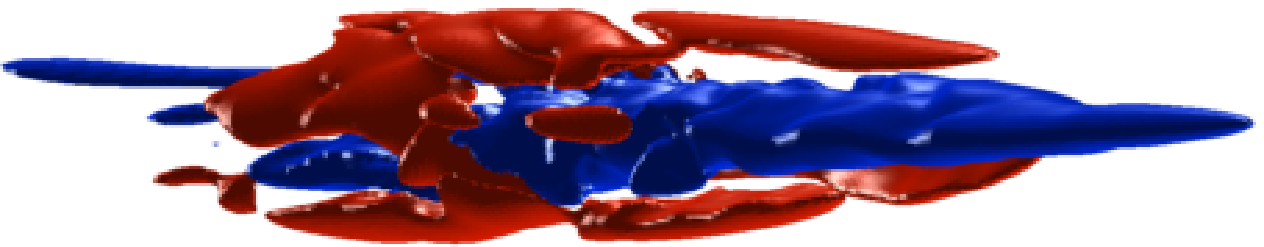}\\ 
$250\,D/U$ & $425\,D/U$ \\ 
\includegraphics[height=0.09\linewidth]{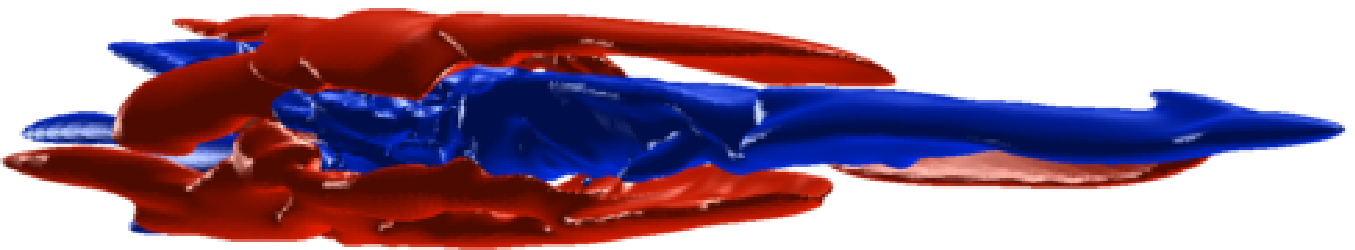}& 
\includegraphics[height=0.09\linewidth]{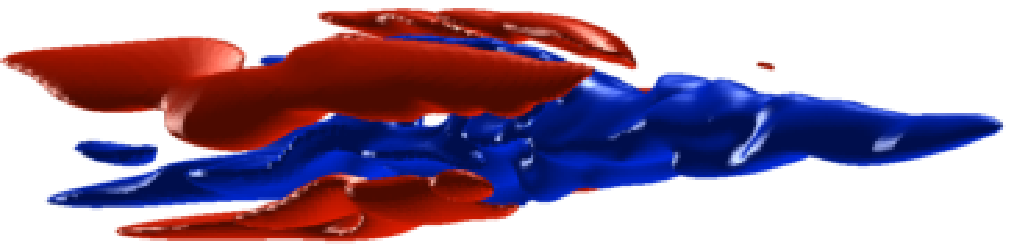} 
\end{tabular} 
\end{tabular} 
\caption{Turbulent puff in pipe flow. 
(a) Time evolution of the pressure gradient required to drive a 
constant flow rate at Reynolds number $\textrm{Re}=1860$, normalised 
with the value for laminar flow. (b) Space-time diagram of the energy 
(logarithmically scaled). The space axis is periodic and corresponds 
to a reference frame moving at the mean speed $U$ of the flow. (c) 
Isosurfaces of streamwise velocity at $0.2\,U$ (red) and $-0.2\,U$ 
(blue). The laminar parabolic profile has been subtracted and the 
axial dimension scaled to aid visualisation.} \label{fig:puff} 
\end{figure}

The importance of spatial interactions in fluid flows can be seen in 
pipes, channels and boundary layers already at low $\textrm{Re}$. In 
these and other wall-bounded shear flows turbulence first appears in 
isolated patches surrounded by laminar 
flow~\cite{reynolds1883,*emmons1951,*tillmark1992,*lemoult2013}. This 
coexistence is possible because the laminar flow is linearly stable 
and finite amplitude disturbances are needed to seed turbulence. In 
pipe flow the Reynolds number (defined as $\textrm{Re}:=DU/\nu$, with 
$D$ the pipe diameter, $U$ the constant mean driving speed and $\nu$ 
the kinematic viscosity of the fluid) is the sole governing 
parameter. Figure~\ref{fig:puff} illustrates the dynamics of a 
localised turbulent patch or puff in a 50-diameter long pipe at 
$\textrm{Re}=1860$. While the mass-flux is kept constant, the driving 
pressure gradient, the streamwise length of the turbulent region and 
its propagation speed strongly fluctuate, and the internal 
arrangement of high and low velocity streaks within the puff evolves 
erratically (see Fig.~\ref{fig:puff}c). These spatio-temporal 
fluctuations result in the eventual collapse of the puff back to 
laminar flow in a memoryless decay 
process~\cite{faisst2004,*hof2006,*peixinho2006}. Although the 
characteristic lifetime of puffs increases with Re, puffs remain 
transient and their dynamics is consistent with a chaotic repeller or 
saddle in the phase space of the Navier--Stokes 
equations~\cite{hof2008,avila2010}. However, as Re is increased, 
puffs can also expand in length and split~\cite{wygnanski1975}, 
thereby progressively filling the pipe with a spatially intermittent 
laminar-turbulent pattern~\cite{moxey2010}. In the limit of long 
pipes, a self-sustained turbulent state first arises once the 
puff-splitting rate exceeds the decay rate~\cite{avila2011}, which 
happens at $\textrm{Re}\approx 2040$. Recently, this transition type 
has been shown to belong to the universality class of directed 
percolation for Couette flow~\cite{lemoult2016}. 

\begin{figure}
\includegraphics[width=\linewidth]{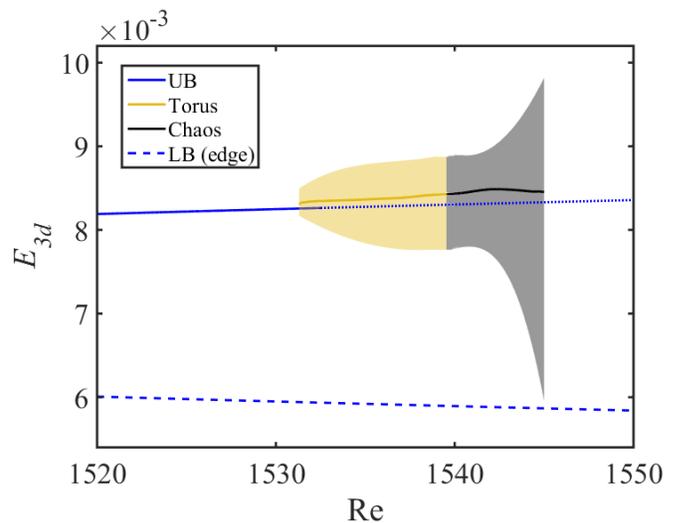}
\caption{Diagram of the time-averaged kinetic energy of the 
  streamwise-dependent part of the velocity field $E_{3d}=||
  \boldsymbol{v}-\langle\boldsymbol{v}\rangle_x||_2^2$ as a function of 
  Reynolds number, depicting the saddle-node bifurcation in 
  \cite{avila2013}. The stable upper branch solution (UB), a
  localised relative periodic orbit, turns into a torus (orange), which 
  breaks up into a chaotic attractor (black). The shaded areas show the 
  variation of energy over long runs. The chaotic fluctuations approach 
  the lower branch solution (LB) separating chaos from the laminar state.}
\label{fig:bifurcations}
\end{figure}

\begin{figure} \centering 
\begin{tabular}{c} 
(a)\\ 
\includegraphics[width=0.744\linewidth]{fig_2a.eps}\\ 
(b)\\ 
\quad\includegraphics[width=0.714\linewidth]{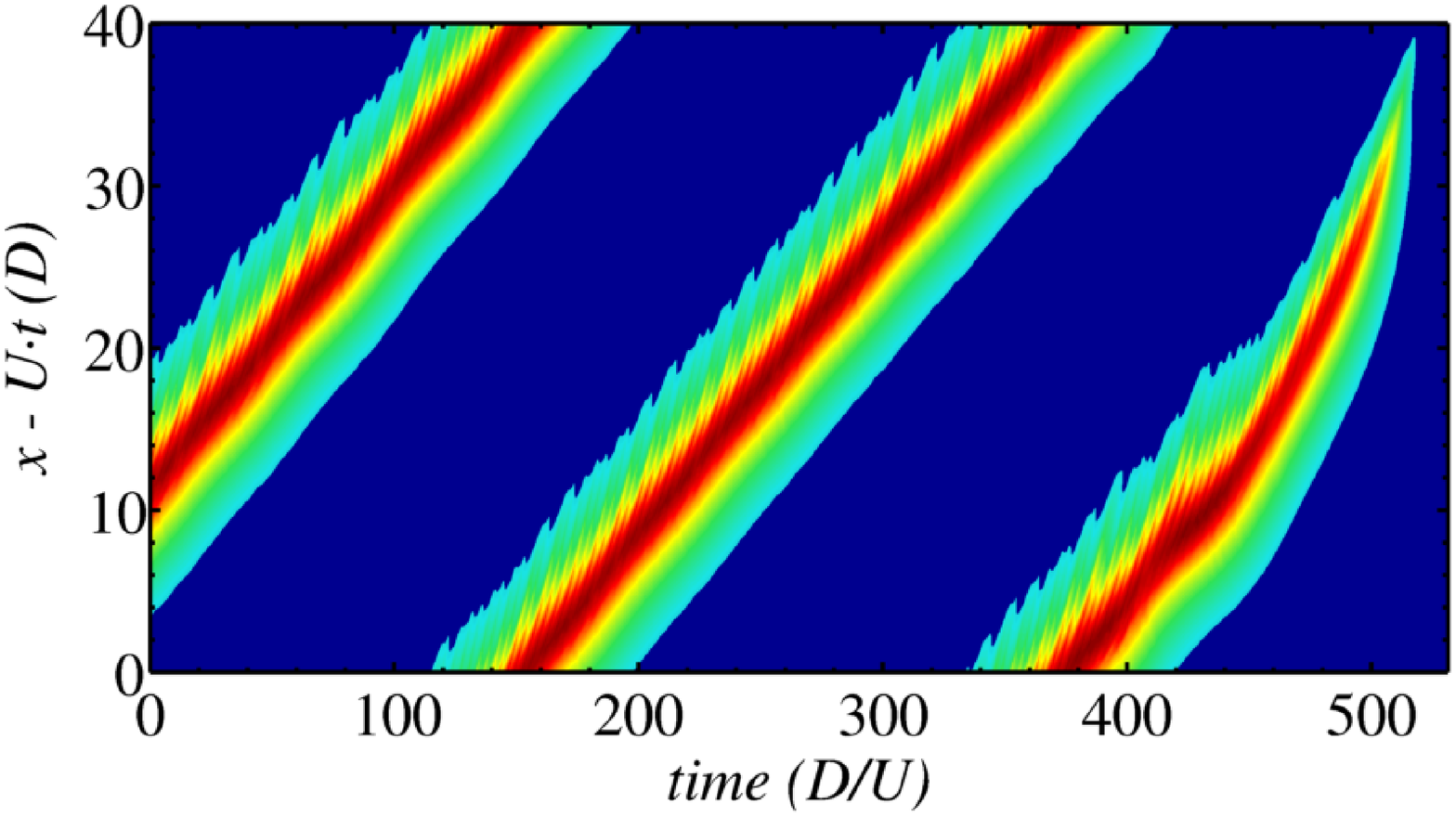}\\ 
\quad(c)\\ 
\begin{tabular}{cc} 
$t=40\,D/U$ & $160\,D/U$ \\ 
\includegraphics[height=0.09\linewidth]{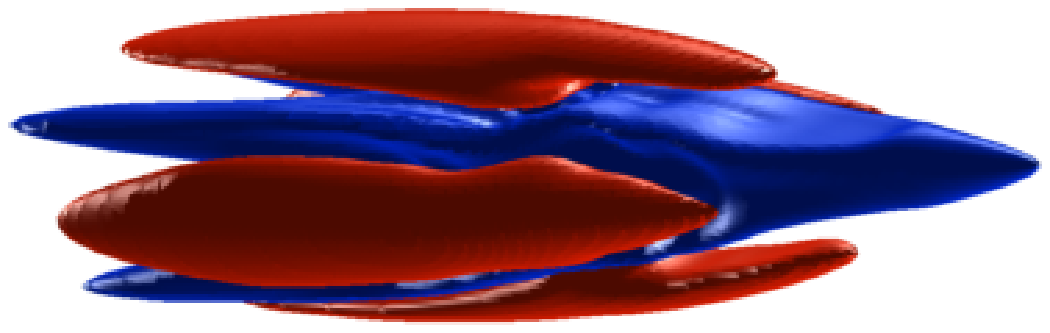}& 
\includegraphics[height=0.09\linewidth]{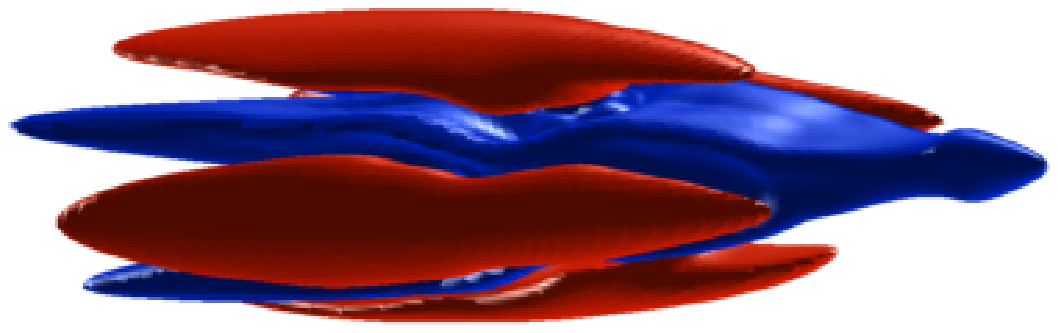}\\ 
$290\,D/U$ & $410\,D/U$\\ 
\includegraphics[height=0.09\linewidth]{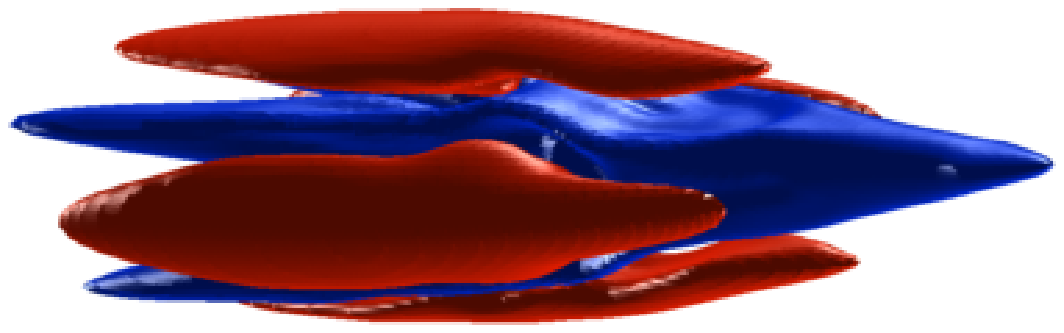}& 
\includegraphics[height=0.09\linewidth]{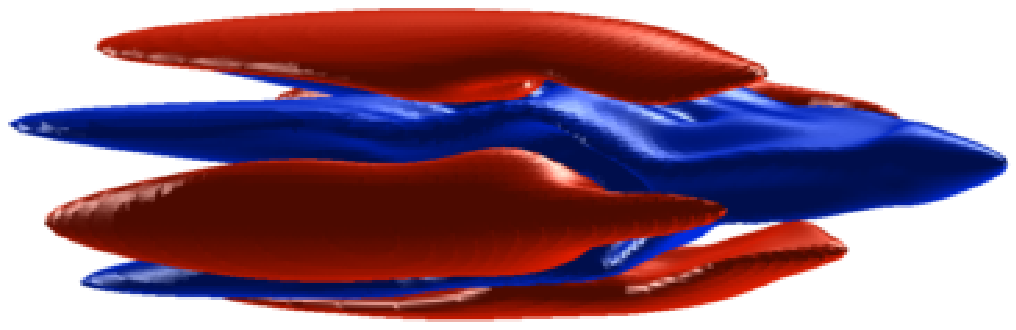} 
\end{tabular} 
\end{tabular} 
\caption{(a) Time evolution of the 
normalised pressure gradient for a chaotic spot at 
$\textrm{Re}=1546$. (b) Corresponding space-time diagram. (c) 
Isosurfaces of streamwise velocity; All axes and colouring as in 
Fig.~\ref{fig:puff}.} \label{fig:chaos} \end{figure}

Despite recent progress in quantifying and 
modelling~\cite{barkley2011,*sipos2011,*allhoff2012,*marschler2014,
*shih2015} the onset of turbulence in pipe flow as a spatio-temporal 
system whose fundamental units are puffs, surprisingly little is 
known about their origin. In fact, a framework providing a plausible 
explanation for their formation just begins to emerge. The key idea 
is that puffs stem from coherent structures (simple invariant 
solutions) of the Navier--Stokes equations that serve as elementary 
building blocks \cite{kerswell2005,*gibson2008,*kawahara2012}. The 
simplest such structures in pipe flow are spatially periodic 
traveling waves \cite{faisst2003,*wedin2004}, which are relative 
equilibria of the governing equations. More recently, streamwise 
localised solutions arising at a saddle-node bifurcation at 
$\textrm{Re}=1428$ have been found~\cite{avila2013} 
(fig.~\ref{fig:bifurcations}). These solutions 
possess reflectional and $\pi$-rotational symmetry, and in this 
subspace the upper branch solution (UB in fig.~\ref{fig:bifurcations}) is 
stable up to $\textrm{Re}=1532.5$, which allows its computation directly by 
time-stepping of the Navier--Stokes equations. The lower branch 
solution (LB in fig.~\ref{fig:bifurcations}) has a single unstable 
direction and is an edge 
state~\cite{skufca2006,*schneider2007} locally separating phase-space 
trajectories that turn chaotic from those that go laminar. At 
$\textrm{Re}\approx 1533$ the upper-branch solution undergoes a subcritical 
Neimark--Sacker bifurcation, with a narrow hysteresis band down to 
$\textrm{Re}=1531.3$, followed by break-up into chaos at 
$\textrm{Re}=1540$. At $\textrm{Re}\gtrsim 1545$ flow trajectories 
were observed to relaminarise, hinting at a global bifurcation 
turning the chaotic attractor into a chaotic saddle or 
repeller~\cite{avila2013} (fig.~\ref{fig:bifurcations}). 
The dynamics of a relaminarising run at 
$\textrm{Re}=1546$ is shown in Fig.~\ref{fig:chaos}. The 
corresponding spatially localised spot exhibits chaotic yet mild 
modulations of structure, streamwise length and propagation speed. 
It is referred to as ``chaotic spot'' in the rest of the
paper in order to make clear that its behaviour is in contrast with 
the strong spatio-temporal fluctuations exhibited by turbulent puffs.
This difference in spatio-temporal complexity, however, raises the
question of what further qualitative changes occur in phase space as
$\textrm{Re}$ increases. 

In this paper we show a mechanism increasing the spatio-temporal 
complexity of chaotic motions in pipe flow. As the Reynolds number is 
increased several chaotic saddles appear in phase space. Each of them 
is associated with localised spots with distinct propagation speed 
and streamwise length. Initially these saddles are rather isolated in 
phase space, but their successive merging results in trajectories 
that explore increasingly larger regions. In physical space, this 
translates into spot length and pressure fluctuations closer to those 
of turbulent puffs. 
 
\section{\textbf Methods} 

In order to render the problem more tractable we simplify the 
dynamics by restricting them subject to reflectional symmetry about a 
diametrical plane and $\pi$-rotational periodicity with respect to the 
pipe axis, as in~\cite{avila2013}. The only exception to this is 
Fig.~\ref{fig:puff}, which was run in full space. Direct numerical 
simulations of the Navier--Stokes equations in cylindrical 
coordinates are carried out using the hybrid spectral 
finite-difference code of Willis~\cite{willis2009}, which is based on 
a velocity-potential formulation automatically satisfying the 
continuity equation and thereby avoiding the computation of the 
pressure. The coupled boundary conditions on the potentials are 
solved to machine precision with the influence-matrix method. We 
refer the interested reader to~\cite{willis2009} for details. 

We use periodic boundary conditions in the streamwise direction, with 
computational domains long enough ($40$ to $100\,D$) such that no 
interaction occurs between the upstream and downstream fronts of the 
turbulent patches. For $\textrm{Re}<1650$ we adopt a time-step 
$\delta t=0.0025\,D/U$, 768 axial Fourier modes ($K=\pm 384$) for a 
length of $50\,D$, 32 azimuthal Fourier modes ($M=\pm 16$) for 
$\theta \in [0,\pi)$ and $N=48$ points in the radial direction. This 
grid resolution and time-step allow us to determine the bifurcation 
sequence leading to transient chaos with a precision better than 
$\Delta\textrm{Re}=\pm 1$. Both time-step and spatial discretisation 
are slightly finer than for previous simulations at larger 
$\textrm{Re}$ in full space~\cite{avila2010} that managed to achieve 
quantitative agreement with turbulent puffs lifetimes measured in 
extremely accurate and precisely controlled 
experiments~\cite{hof2008}. For $\textrm{Re}\ge 1650$ we use a domain 
length of $100\,D$ and $K=\pm 1024$ to accurately resolve the 
increasingly turbulent dynamics. 
 
\section{Boundary crisis to transient chaos}\label{sec:bc} 

\begin{figure}
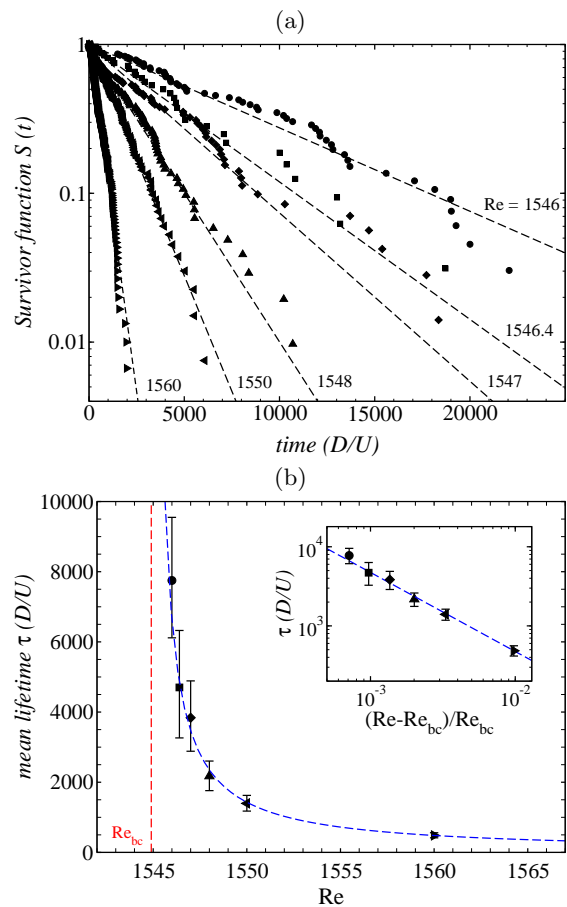

  \centering
  \begin{tabular}{c}
    (a)\\
    \includegraphics[width=0.85\linewidth]{fig_3a.eps}\\
    (b)\\
  \includegraphics[width=0.85\linewidth]{fig_3b.eps}
  \end{tabular}
  \caption{(a) Survival probability of chaotic transients as a function of 
  time for $\textrm{Re}\, \in\, [1546,1560]$. Initial conditions were taken from 
  a chaotic spot at $\textrm{Re}=1545$. (b) Mean lifetime $\tau$ extracted 
  from the distributions in (a), with the corresponding confidence 
  intervals estimated as in \cite{avila2011}. The vertical (red) dashed 
  line shows the critical Reynolds number $\textrm{Re}_\text{bc} = 1545$. 
  This was obtained with an inverse fit to equation~\eqref{eq:tau_bc}, 
  which is shown as a (blue) dashed line. The inset shows the lifetime 
  scaling as function of reduced Reynolds number in log-log axes.}
  \label{fig:lifetime}
\end{figure}

We start by analysing the transition from persistent to transient 
chaotic spots, which was previously estimated to occur at 
$\textrm{Re} \gtrsim 1545$~\cite{avila2013}. In order to shed light 
on the underlying bifurcation mechanism, we perform a statistical 
study of the mean lifetime ($\tau$) of the transients. Initial 
conditions for this study were taken from a chaotic spot at 
$\textrm{Re}=1545$. Figure~\ref{fig:lifetime}a shows the survival 
probability of the transients as a function of elapsed time for 
several $\textrm{Re}>1545$. As in low-dimensional systems 
relaminarisation follows a memoryless decay characterised by the 
escape rate $1/\tau$ from a chaotic saddle~\cite{tel2008,*tel2015}. 
As $\textrm{Re}$ increases, the mean lifetime rapidly decreases and 
is well described by 
\begin{equation} 
\tau=\dfrac{a}{\textrm{Re}-\textrm{Re}_\text{bc}},\label{eq:tau_bc} 
\end{equation} 
with critical Reynolds number 
$\textrm{Re}_\text{bc}=1545$ and $a=7293\,D/U$ obtained from an 
inverse linear fit (blue dashed line in~Fig.~\ref{fig:lifetime}b and 
inset). Note that in experiments the lifetimes of turbulent puffs 
increase super-exponentially for $\textrm{Re}>1700$~\cite{hof2006}, 
which suggests that further qualitative changes must occur in phase 
space as Re increases. The simple scaling of 
equation~\eqref{eq:tau_bc} is typical of low-dimensional chaos 
\cite{tel2008}, and suggests that the onset of transients may take 
place via a generic mechanism called boundary crisis 
\cite{grebogi1987}. In a boundary crisis, the attractor collides with 
its own basin boundary and opens up to become a strange or chaotic 
saddle \cite{tel2008}. 

\begin{figure}
  \centering
  \includegraphics[width=0.87\linewidth]{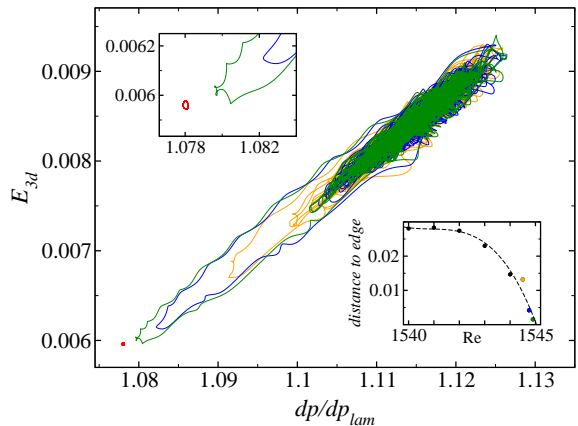}
  \caption{Phase-portrait of the chaotic attractor at $\textrm{Re}=1544.5$ 
  (orange), $1544.75$ (blue) and $1544.9$ (green). The dynamics are 
  projected onto a two-dimensional plane defined by the normalised pressure 
  gradient $dp/dp_{lam}$ and the kinetic energy of the streamwise-dependent 
  part of the velocity field $E_{3d}$. The trajectories 
  approach the ``edge state'' on the basin boundary of the attractor, which 
  is shown in red. The upper-left inset shows a closeup near the edge 
  state. The bottom right inset shows how the distance between the 
  attractor and its basin boundary (as defined by minimum pressure gradient 
  difference with the edge state over time) goes to zero as 
  $\textrm{Re}_\text{bc} = 1545$ is approached. The dashed line helps 
  to guide the eye (cubic fit to the data).}
  \label{fig:bc}
\end{figure}

This hypothesis is put to test here by examining the behaviour of 
extremely long phase-space trajectories ($> 10^5\,D/U$) of the 
attractor in the range $\textrm{Re}\in[1540,1545]$. Fig.~\ref{fig:bc} 
shows that the chaotic trajectories sporadically visit the 
neighbourhood of the lower branch solution, which is known to be the 
attractor at the basin boundary (or edge 
state~\cite{skufca2006,*schneider2007}) in the 
subspace~\cite{avila2013}. The edge state is visited more closely 
with increasing Re and, as $\textrm{Re}_\text{bc}$ is approached, the 
minimum distance to the edge state vanishes (bottom-right inset of 
Figure~\ref{fig:bc}). This trend supports the boundary crisis 
hypothesis and a cubic fit to the data yields $\textrm{Re}_\text{bc} 
= 1545$, exactly as for the lifetimes analysis. 
Crises leading to chaotic transients were recently observed in 
Couette~\cite{kreilos2012,*shimizu2014}, channel~\cite{zammert2015} 
and pipe flows~\cite{altmeyer2015}, highlighting their ubiquity in 
shear flows. Note however that all these studies were done in short 
periodic domains unable to capture spatial localisation, a salient 
feature of shear flow turbulence. 

\section{Onset of spatio-temporal fluctuations via saddle 
merger}\label{sec:merger} 

\begin{figure} \centering \begin{tabular}{c} (a)\\ 
\includegraphics[width=0.75\linewidth,clip=]{fig_5a.eps}\\ 
(b)\\ 
\quad\includegraphics[width=0.714\linewidth,clip=]{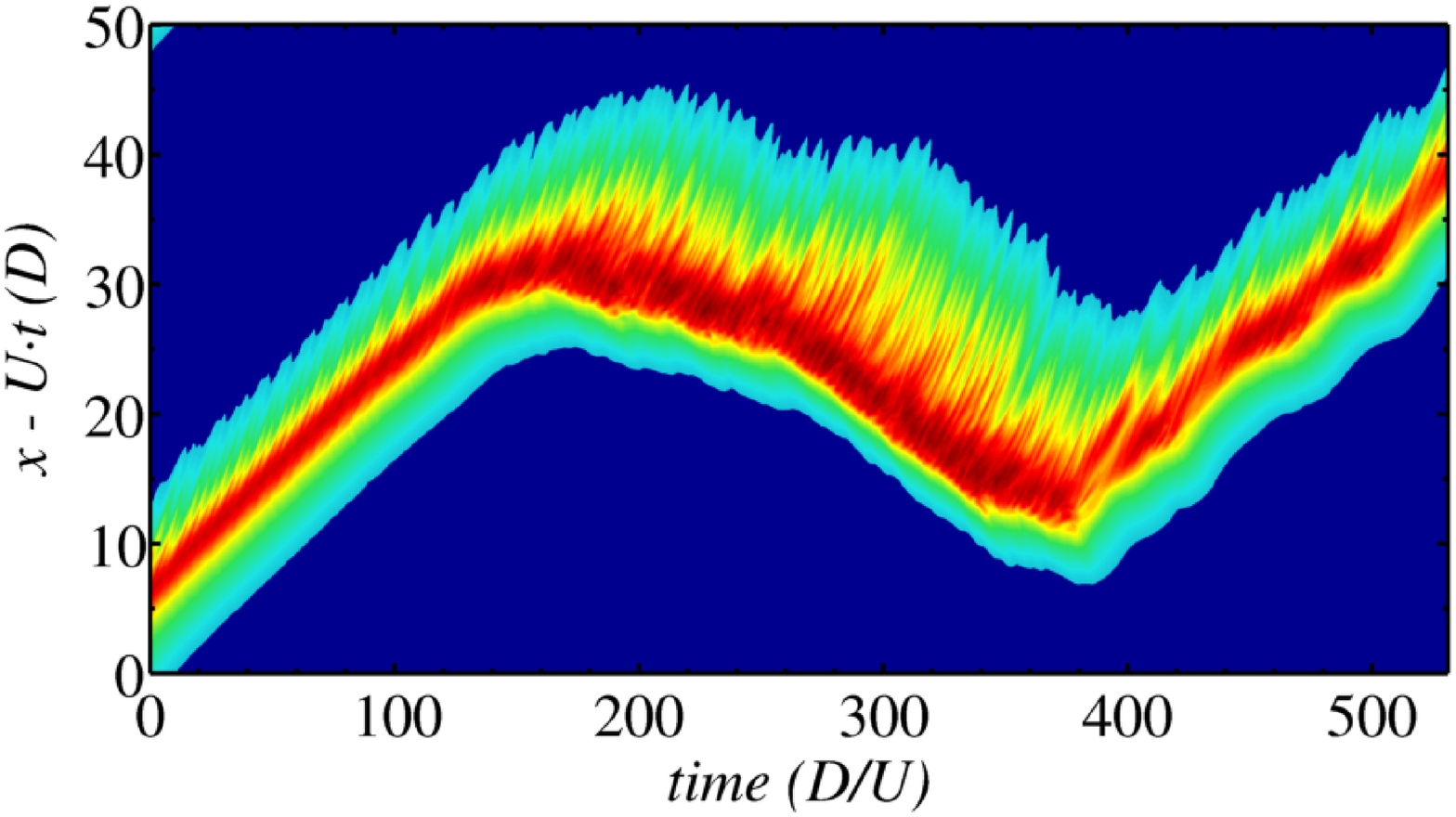}\\ 
(c)\\ \begin{tabular}{cccc} $t=60\,D/U$ & $200\,D/U$ \\ 
\includegraphics[height=0.09\linewidth]{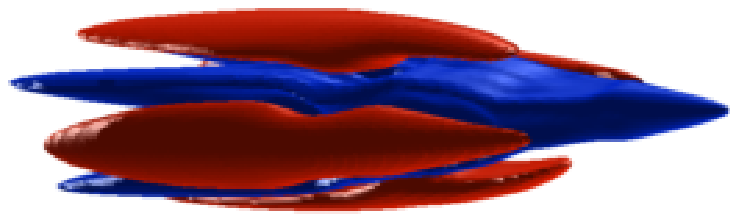}& 
\includegraphics[height=0.09\linewidth]{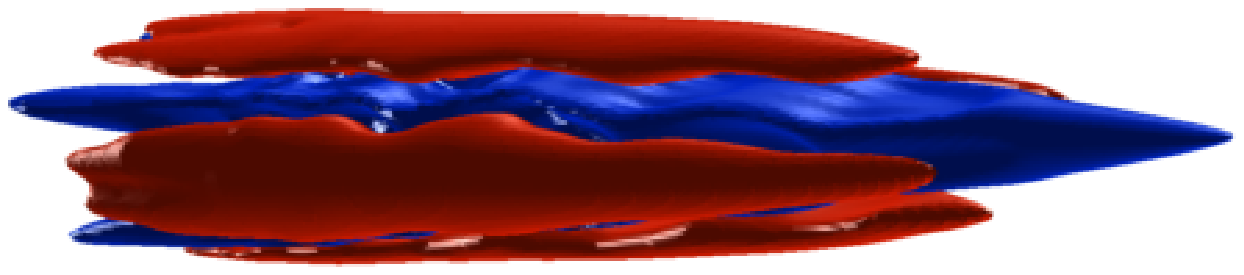}\\ 
$300\,D/U$ & $460\,D/U$ \\ 
\includegraphics[height=0.09\linewidth]{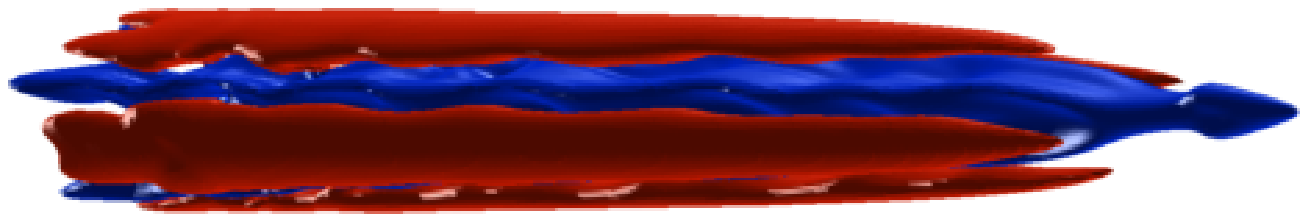}& 
\includegraphics[height=0.09\linewidth]{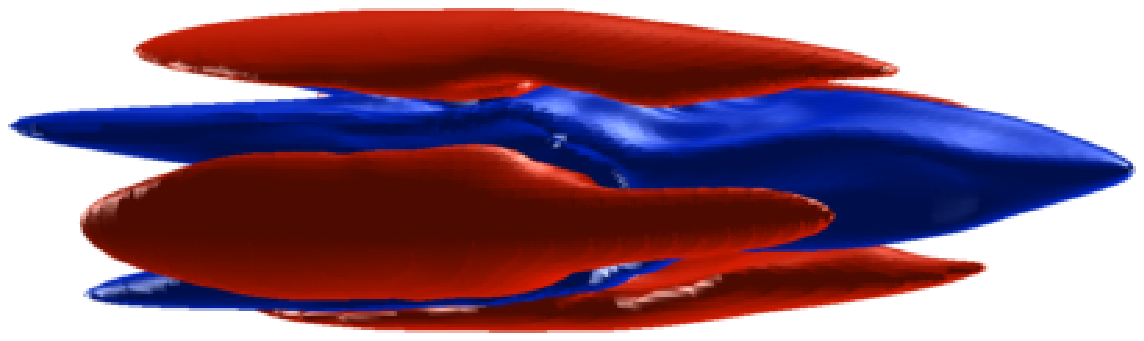} 
\end{tabular} \end{tabular} \caption{Emergence of turbulent 
dynamics from interacting chaotic spots. At $\textrm{Re}=1580$ the 
flow exhibits sudden changes in dynamical and kinematic properties 
before relaminarising. These occur through the interaction of two 
chaotic saddles corresponding to short (A) and long (B) chaotic 
spots. (a) Excerpt of a trajectory. (b) Corresponding space-time 
diagram. (c) Isosurfaces of streamwise velocity; All axes and 
coloring as in Fig.~\ref{fig:puff}.} \label{fig:complexity} 
\end{figure}

The mild fluctuations displayed by the transient chaotic spot 
together with the simple lifetime scaling suggest that the dynamics 
involved is low-dimensional. However, traces of more complex 
behaviour can be observed at larger $\textrm{Re}$. 
Figure~\ref{fig:complexity}a shows the time evolution of the driving 
pressure gradient corresponding to a run at $\textrm{Re}=1580$, which 
was initialised from a chaotic spot at $\textrm{Re}=1545$ and
relaminarises at a later time (not shown). Initially 
the dynamics remains on a low friction plateau (labelled ``A''), and 
the spot has a streamwise length and propagation speed similar to 
that observed at lower $\textrm{Re}$. This is evidenced by comparison 
of the space-time diagram of Fig.~\ref{fig:complexity}b up to 
$t=150\,D/U$, to its counterpart at $\textrm{Re}=1546$ 
(Fig.~\ref{fig:chaos}b). Subsequently the dynamics explores a region 
of higher friction (labelled ``B''), which corresponds to a spatial 
expansion of the spot (see the snapshots of 
Fig.~\ref{fig:complexity}c). Thereafter the spot rapidly recedes in 
length and the dynamics settles anew on the low friction plateau A. 

\begin{figure}
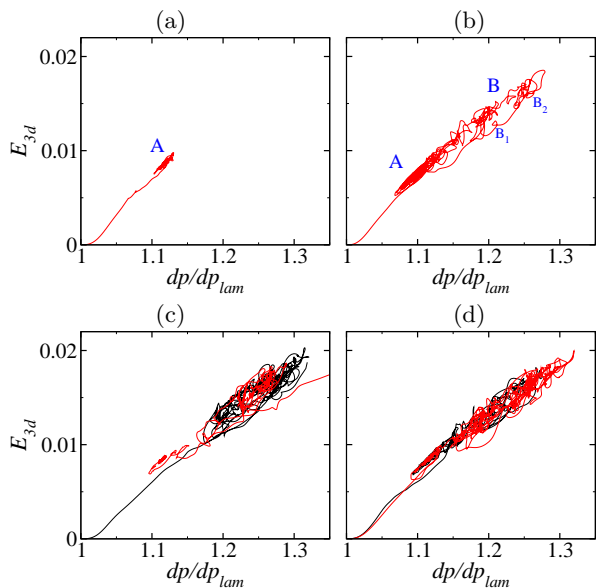
 \centering \begin{tabular}{cc} (a) & (b) \\ 
\includegraphics[scale=0.23,clip=]{fig_6a.eps} & 
\includegraphics[scale=0.23,clip=]{fig_6b.eps} \\ (c) & (d) 
\\ \includegraphics[scale=0.23,clip=]{fig_6c.eps} & 
\includegraphics[scale=0.23,clip=]{fig_6d.eps} \end{tabular} 
\caption{Phase portraits in the vicinity of the boundary crisis 
($\textrm{Re}_\text{bc} \approx 1545$). (a) Dynamics slightly after 
the crisis ($\textrm{Re}=1546$, run of Fig.~\ref{fig:chaos}) and (b) 
far beyond the crisis ($\textrm{Re}=1580$, run of 
Fig.~\ref{fig:complexity}). Initial conditions for (a,b) are taken 
from a chaotic spot at $\textrm{Re}=1545$. (c) Dynamics prior to the 
crisis ($\textrm{Re}=1540$): one trajectory starts in region B and 
evolves towards A where it stays forever (red) while the other 
relaminarises directly from region B (black). (d) Dynamics after the 
crisis ($\textrm{Re}=1550$): one trajectory relaminarises from region 
A after having visited B (red) whereas the other relaminarises from B 
after having visited A (black). Initial conditions for (c,d) are 
taken from a turbulent puff at $\textrm{Re}=2000$. } 
\label{fig:pmaps} \end{figure}

The phase portraits of Fig.~\ref{fig:pmaps} illustrate the dynamics 
of the flow for several $\textrm{Re}$. At $\textrm{Re}=1546$ the 
dynamics revolves around a fairly small region of phase space (A), 
whereas a much larger region is explored at $\textrm{Re}=1580$. This 
qualitative change suggests that two separate regions of phase space, 
(A) and (B), originally dynamically isolated, have merged at a global 
bifurcation. The chaotic spot that results from the merger exhibits 
strong fluctuations in its propagation speed and spatial extent (see 
Fig.~\ref{fig:complexity}b--c)). To elucidate the nature of the 
underlying global bifurcation we perform an additional set of runs 
with initial conditions taken from a turbulent flow at larger 
$\textrm{Re}=2000$, thus ensuring that the dynamics are not biased to 
start from too close to A. 

\subsection{Dynamics prior to the boundary 
crisis}\label{sec:before_crisis} 

\begin{figure}
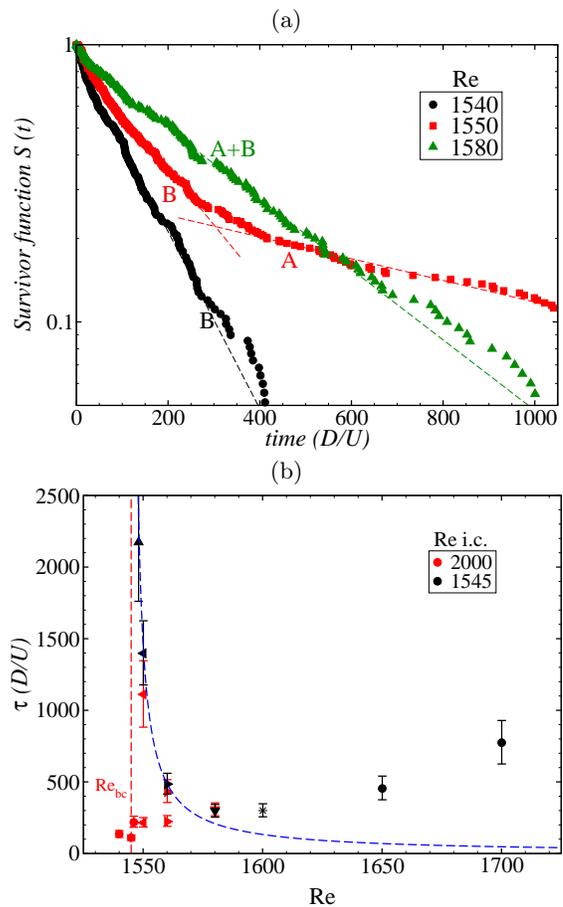
 \centering \begin{tabular}{c} (a)\\ 
\includegraphics[width=0.84\linewidth,clip=]{fig_7a.eps}\\ 
(b)\\ 
\includegraphics[width=0.85\linewidth,clip=]{fig_7b.eps} 
\end{tabular} \caption{(a) Survival probability obtained from 
initial conditions at $\textrm{Re}=2000$. The data at 
$\textrm{Re}=1550$ (red) exhibits two slopes, which reflect the 
interaction of two merging chaotic saddles (A and B). (b) Mean 
lifetime extracted using initial conditions from either the chaotic 
saddle A at $\textrm{Re}=1545$ (black, same data as in 
Fig.~\ref{fig:lifetime}b) or turbulent flow at $\textrm{Re}=2000$ 
(red, see distributions in (a)).} \label{fig:merger} \end{figure}

We first reduce the Reynolds number from $\textrm{Re}=2000$ to 
$\textrm{Re}=1540$, which precedes the boundary crisis 
($\textrm{Re}_\text{bc}=1545$). Here A is an attractor and 
phase-space trajectories evolving towards it remain there forever 
(see the red curve in Fig.~\ref{fig:pmaps}c). Other trajectories 
approach region B instead before finally relaminarising (black 
curve). Of a total of 300 such runs, 22\% converged onto A (either 
directly or after an interim visit to B) while the remaining 78\% 
visited B before relaminarising. The survivor function of the latter 
is marked with black symbols in Fig.~\ref{fig:merger}a, and its 
exponential trend confirms the existence of a second saddle B. 

Figure~\ref{fig:pmaps}c shows that B holds transient dynamics that 
eventually leave towards either laminar flow (black curve) or 
coexisting attractor A (red curve). Following the boundary crisis, A 
also becomes a saddle and all trajectories ultimately relaminarise 
regardless of whether they temporarily settled on either A or B 
(Fig.~\ref{fig:pmaps}d). 

\subsection{Weak saddle interaction after the crisis} 

At slightly larger $\textrm{Re}=1550$ trajectories occasionally 
switch between the two saddles (Fig.~\ref{fig:pmaps}d), which 
supports that saddles A and B have undergone a merging 
crisis~\cite{feng2011}. Visits to both A and B are rare and two 
slopes, corresponding to either saddle, are clearly discernible in 
the survivor function for $\textrm{Re}=1550$ (red symbols in 
Fig.~\ref{fig:merger}a). The points on the right slope correspond to 
trajectories that evolve towards A and have longer lifetimes than 
those that approach B (left slope). Note that pinpointing the precise 
$\textrm{Re}$ at which the merger takes place would require 
substantially longer runs and finer $\textrm{Re}$-steps than are 
feasible in reasonable time scales. However, our data suggests that 
the merger occurs very closely, if not immediately, after the 
boundary crisis. The combined saddle features enhanced 
spatio-temporal chaos. 

\subsection{Progressive merger of saddles} 

\begin{figure} \centering \begin{tabular}{cc} (a) & (b) \\ 
\includegraphics[scale=0.15,clip=]{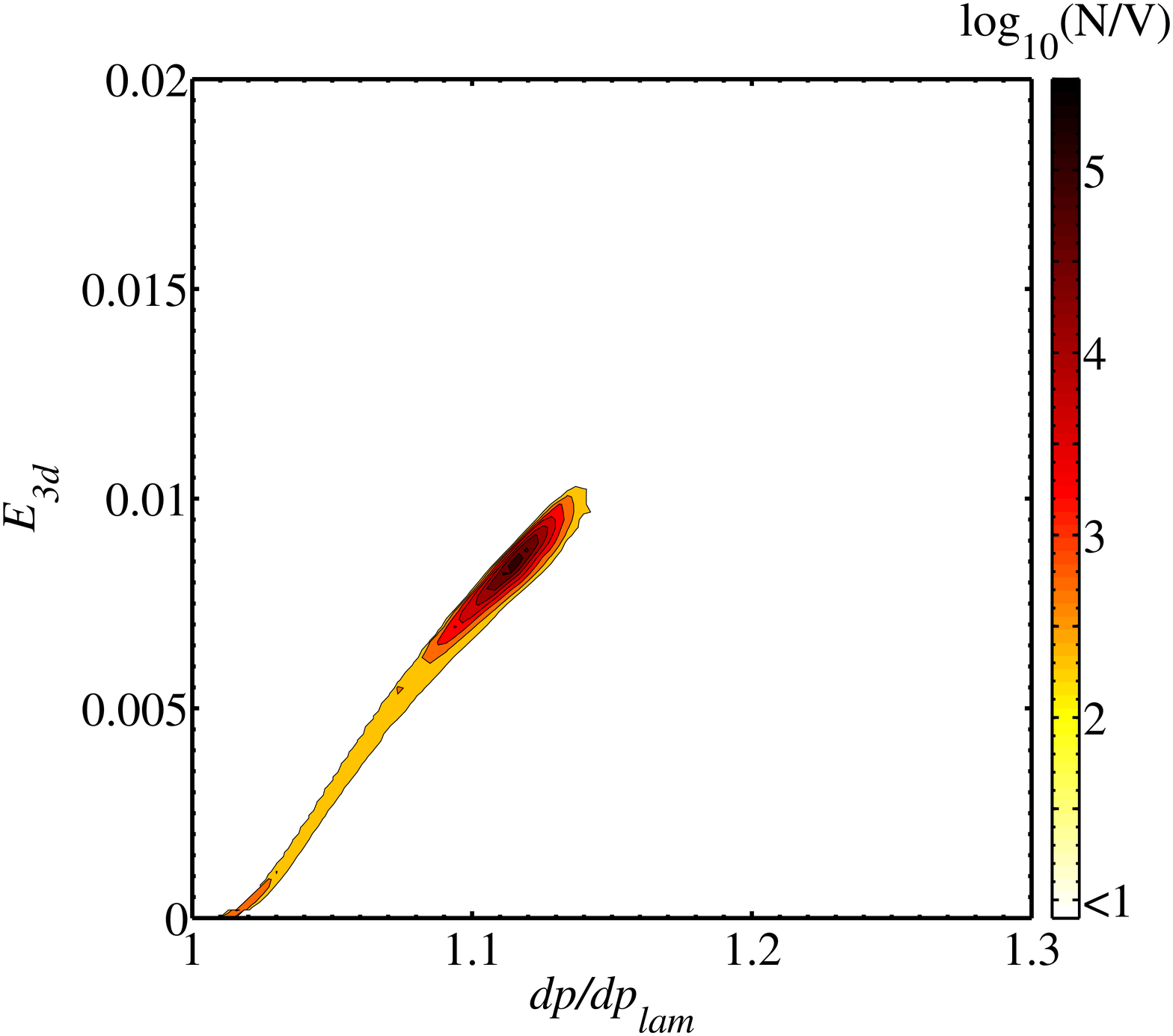} & 
\includegraphics[scale=0.15,clip=]{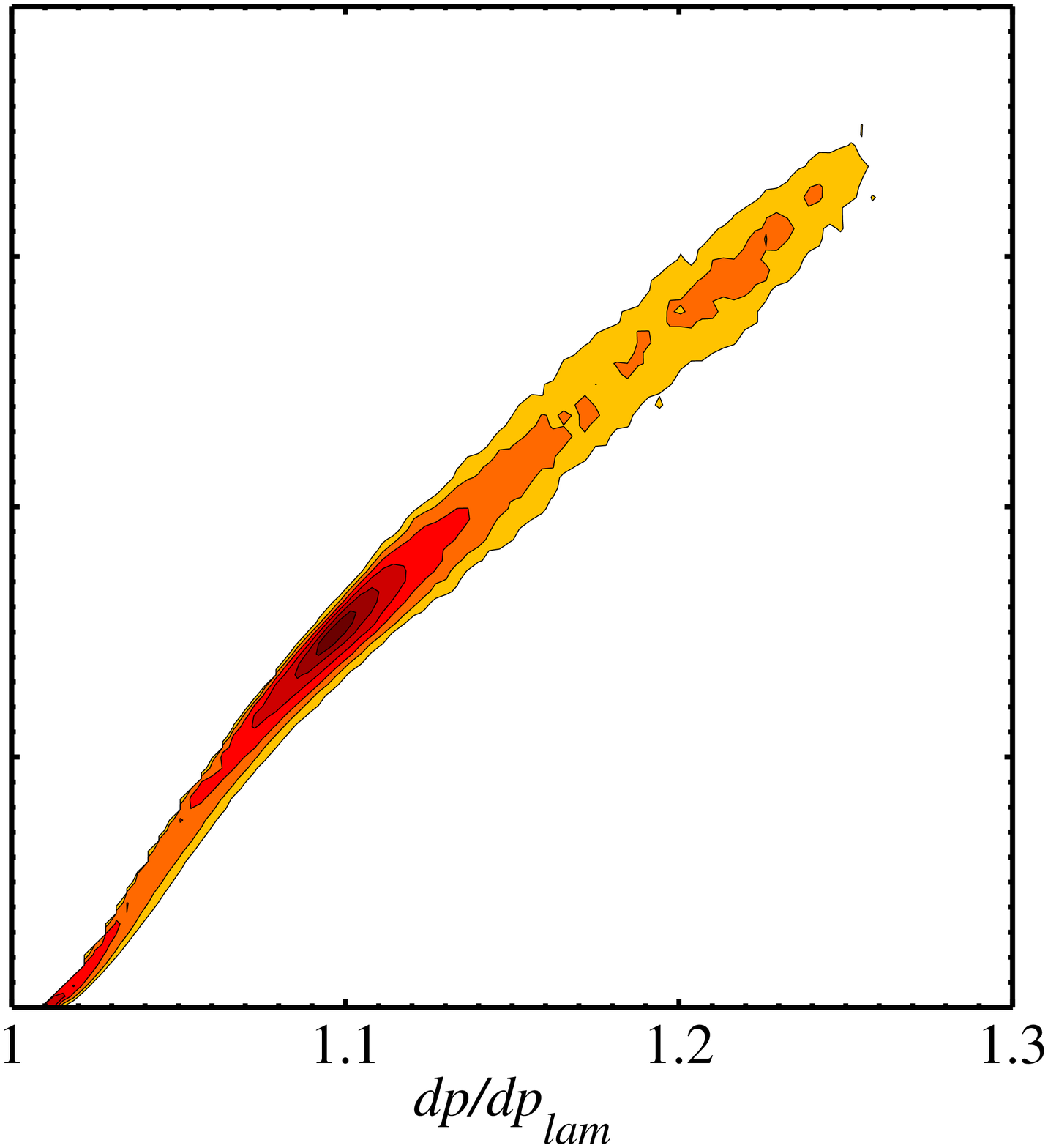} \\ (c) & (d) 
\\ 
\hspace{-5.7mm}\includegraphics[scale=0.15,clip=]{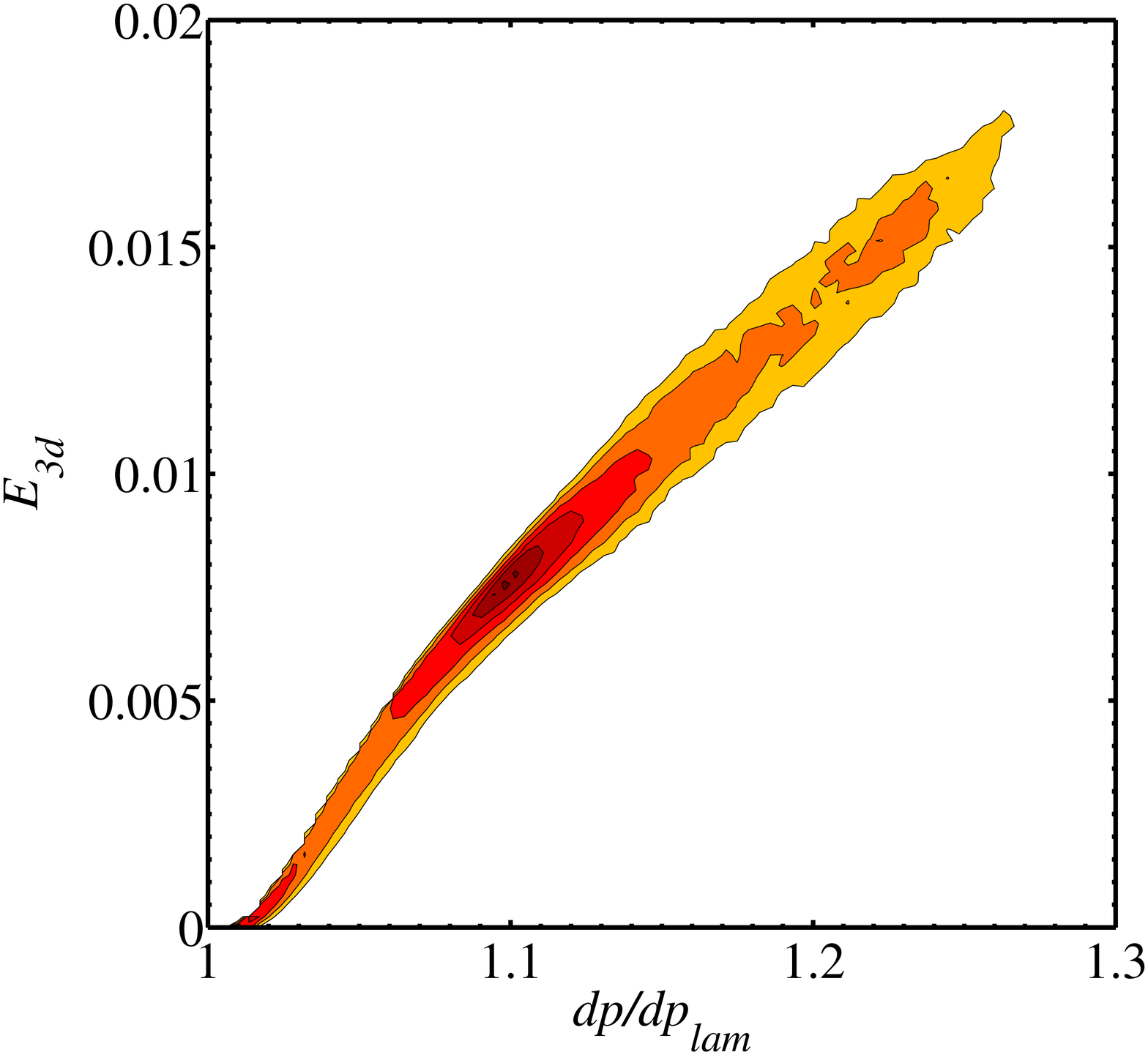} & 
\includegraphics[scale=0.15,clip=]{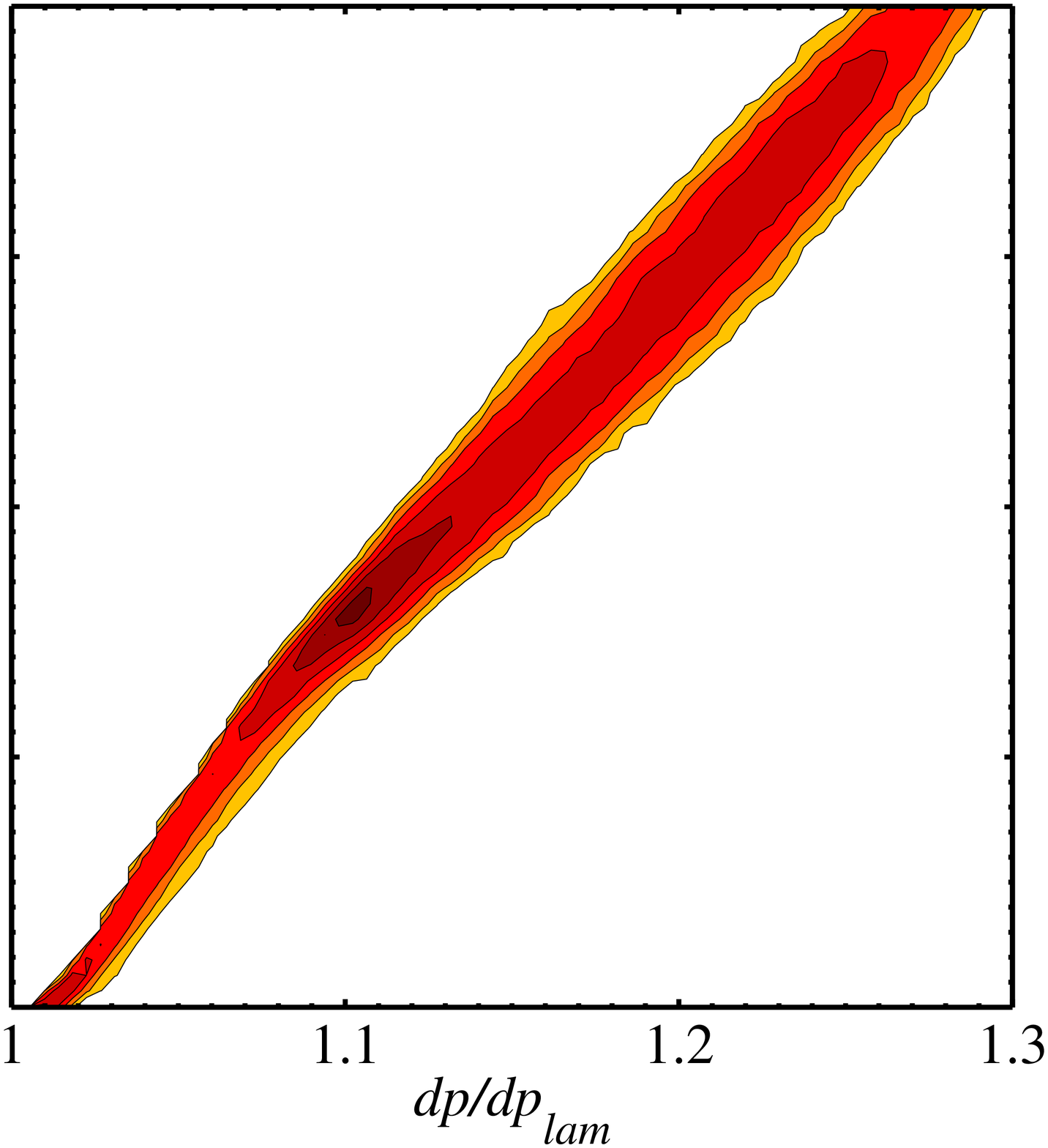} \end{tabular} 
\caption{Phase space histograms (logarithmically scaled) of all runs 
at various Re with initial conditions at $\textrm{Re}=1545$. $N/V$ is 
the count normalized with the volume of the histogram. 
(a)~$\textrm{Re}=1550$. (b) $\textrm{Re}=1580$. (c) 
$\textrm{Re}=1600$. (d) $\textrm{Re}=1650$.} \label{fig:pdfs} 
\end{figure}

\begin{figure} \centering \begin{tabular}{c} (a)\\ 
\includegraphics[width=0.84\linewidth,clip=]{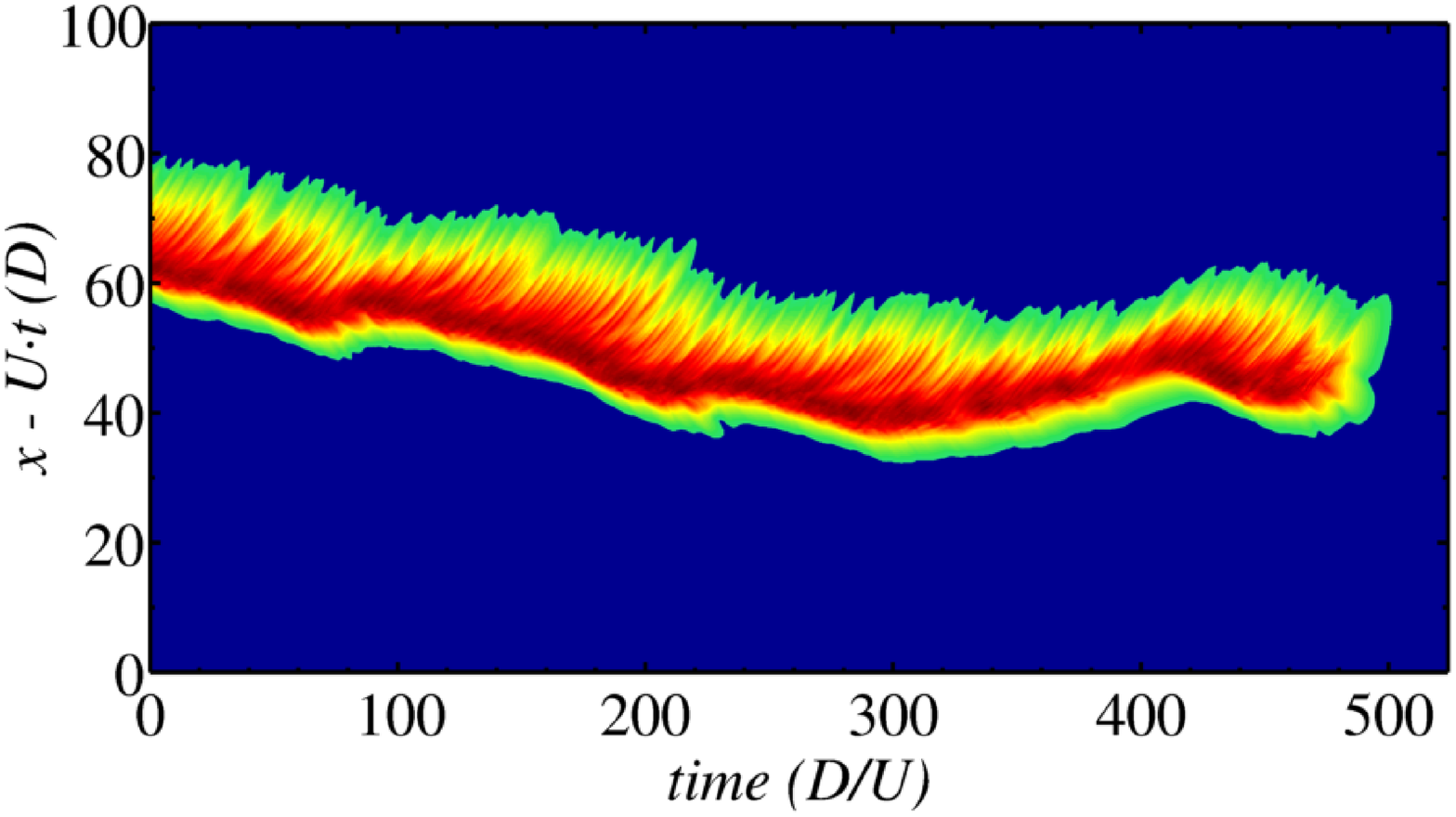}\\ 
(b)\vspace{1mm}\\ \begin{tabular}{cc} $t=80\,D/U$ & 
$t=215\,D/U$ \\ 
\includegraphics[scale=0.1,clip=]{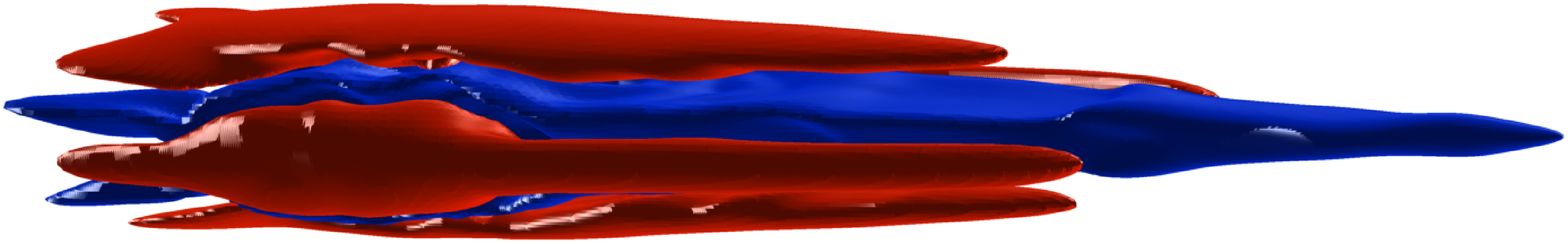} & 
\includegraphics[scale=0.1,clip=]{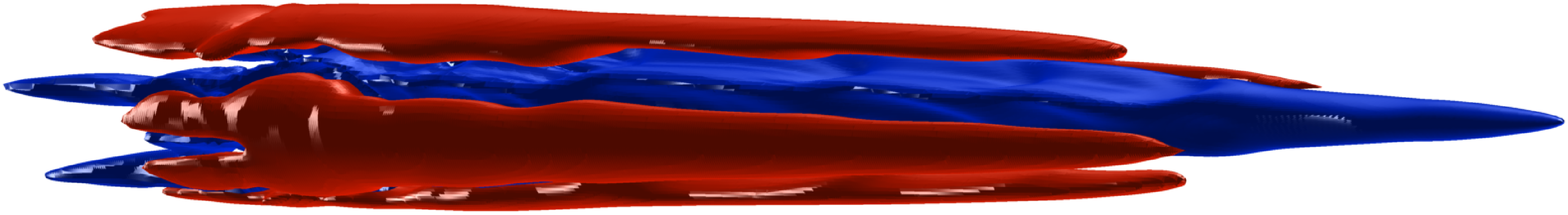} \\ 
$t=340\,D/U$ & $t=465\,D/U$ \\ 
\includegraphics[scale=0.1,clip=]{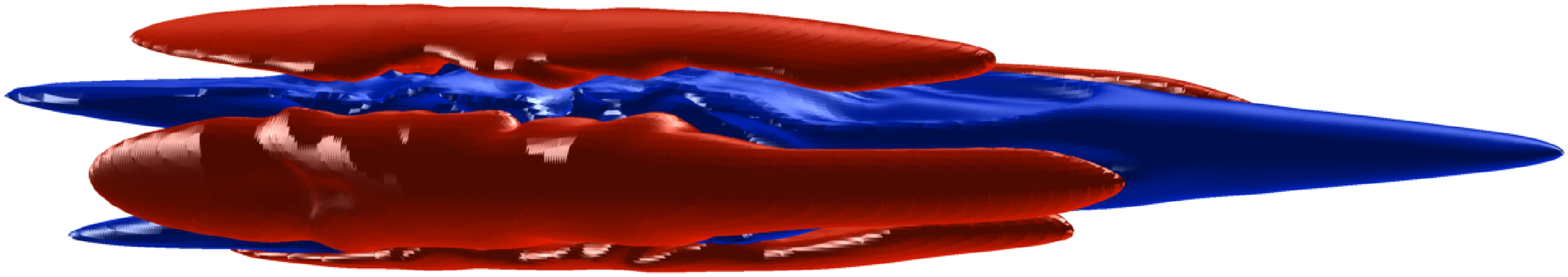} & 
\includegraphics[scale=0.1,clip=]{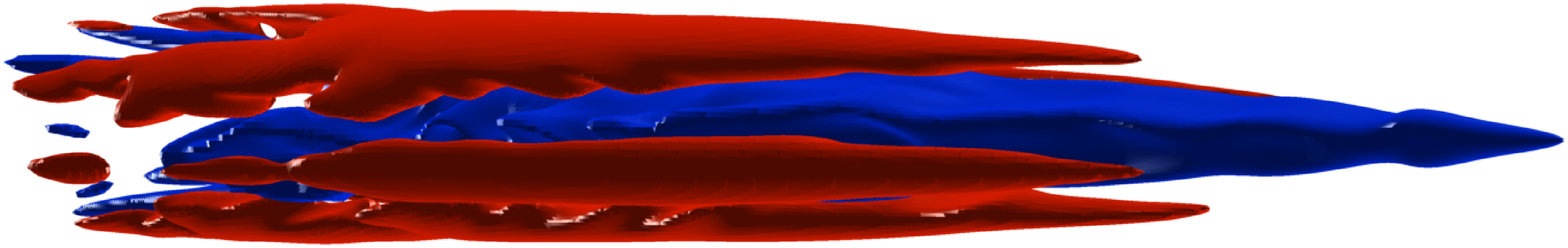}\\ 
\end{tabular} \end{tabular} \caption{(a) Space-time diagram 
and (b) streamwise velocity isosurfaces at $\textrm{Re}=1700$; Axes 
and colouring as in Fig.~\ref{fig:puff}.} \label{fig:spt} 
\end{figure}

As $\textrm{Re}$ is further increased the ghosts of the two saddles 
progressively blend into one sole entity (Fig.~\ref{fig:pdfs}). 
At $\textrm{Re} \approx 1580$ the characteristic times 
for transitions between A and B become shorter than their 
relaminarisation times, so most trajectories visit both A and B 
several times before relaminarising (see Figs.~\ref{fig:complexity} 
and \ref{fig:pmaps}b). Henceforth the two saddles effectively become 
one and most phase-space trajectories explore the total phase space 
associated beforehand with A and B (see Figs.~\ref{fig:pmaps}b, 
\ref{fig:pmaps}d and \ref{fig:pdfs}b). As a consequence, a unique 
combined slope emerges in the survival function (green symbols in 
Fig.~\ref{fig:merger}a). As the transition rate among saddles 
increases, the clearly identifiable regimes of 
Fig.~\ref{fig:complexity} gradually transform into the puff-like 
dynamics of Figs.~\ref{fig:spt}a--b. The merger in phase space is 
accompanied by a rise in lifetimes with increasing $\textrm{Re}$ 
(Fig.~\ref{fig:merger}b) as trajectories bounce from saddle to saddle 
more and more frequently. This reverses the initial decreasing trend 
followed after the boundary crisis and is thus consistent with 
experimental observations~\cite{hof2006,hof2008}. 

Kreilos \emph{et al}.~\cite{Kreilos2014} have also reported a 
mechanism increasing the lifetimes of chaotic transients. In their 
simulations of minimal-flow-unit plane Couette flow an attractor 
arises within an existing saddle. The attracting bubble bursts in a 
crisis and the new saddle features its own lifetime statistics, which 
differ from those of the pre-existing saddle. As a result, the 
combined ensuing lifetimes increase in a non-smooth and non-monotonic 
fashion, which is in stark contrast to the monotonous increase of 
lifetimes that follows the saddle merger observed here. 

\subsection{Bifurcation mechanisms behind saddle merger} 

In the last decade several PDE-based models with at least one spatial 
extended dimension have been employed to illustrate the behaviour of 
chaotic saddles in high-dimensional dynamical 
systems~\cite{rempel2003,*rempel2004}. Particular attention has been 
paid to the type of crises leading from temporal to spatio-temporal 
chaos~\cite{rempel2007,*rempel2007b,tel2008}. Despite the important 
role of chaotic saddles in these models, their ensuing 
spatio-temporal chaos is always attracting. The underlying attractor 
results from an interior crisis of a temporally chaotic attractor 
evolved from a saddle node bifurcation. The spatio-temporal dynamics 
are already encoded in the surrounding chaotic saddle within which 
the interior crisis occurs, but since the only leak of the saddle is 
towards the embedded attractor, the resulting set is an enlarged 
spatio-temporally chaotic attractor that encompasses the pre-existing 
attractor and saddle. The scenario changes dramatically whenever the 
basin boundary separating laminar flow from the second attractor has 
previously undergone a smooth-fractal and, possibly, subsequent 
fractal-fractal metamorphoses~\citep{grebogi1986b,*grebogi1987b}. 
Then a direct transition from permanent temporal chaos to transient 
spatio-temporal chaos becomes feasible~\cite{chian2013}. 

A general theory relating crises of chaotic sets, basin boundary 
metamorphoses and explosions of chaotic saddles~\citep{robert2000} 
sets the framework in which our results may be interpreted. Upon the 
genesis of saddle A in a boundary crisis no permanent dynamics 
remain. The explosion of attractor A into saddle A+B and the 
pre-existence of saddle B suggest that the crisis involves a fractal 
basin boundary, much as in \cite{rempel2007,*rempel2007b,rempel2009}, 
except that the resulting spatio-temporal chaotic set is a saddle 
instead of an attractor. The fact that spatio-temporally chaotic 
saddle B can drive trajectories indistinctly towards A or laminar 
flow, indicates that it is embedded in the basin boundary, which is 
fractal. As the boundary crisis occurs and temporal chaotic attractor 
A becomes a saddle, A is subsumed into an enlarged spatio-temporal 
chaotic saddle A+B within the basin boundary. At low $\textrm{Re}$ 
the basin boundary is smooth and presumably becomes fractal in a 
smooth-fractal metamorphosis~\cite{grebogi1986b}. This metamorphosis 
might be related to the very creation of saddle B, which would have 
emerged within the basin boundary or fused with it in its own crisis. 
Note that unlike for shear flows none of the aforementioned studies 
on PDE-based models involves localised structures. 
 
\section{Origin of chaotic saddles} 

In \S\ref{sec:bc} we have demonstrated that saddle A emerges as the 
chaotic attractor grows and finally collides with its basin boundary 
at $\textrm{Re}_\text{bc}=1545$. This raises the question of whether 
saddle B results from a similar global bifurcation. Note however that 
the dynamics of B is much more complex than that of A: in 
Figure~\ref{fig:complexity}b two distinct friction plateaus (B$_1$ 
and B$_2$) are clearly discernible within B, and the phase portrait 
of Fig.~\ref{fig:pmaps}b also shows that the dynamics of B$_1$ and 
B$_2$ unfolds in two distinct regions of phase space. The space-time 
diagram of Fig.~\ref{fig:complexity}b and snapshots of streamwise 
velocity isosurfaces in Fig.~\ref{fig:complexity}b--c further 
evidence, now in physical space, that B$_1$ and B$_2$ correspond to 
two spots with different streamwise length and propagation speed. 

In an attempt to elucidate the origin of saddle B, we performed an 
extensive study of its lifetimes statistics. Interestingly, at 
$\textrm{Re}\sim 1470$ lifetime distributions feature two clear 
slopes corresponding to spots B$_1$ and B$_2$. These are similar to 
the lifetime distribution shown as red in Fig.~\ref{fig:merger}a and 
hence not shown here. For $\textrm{Re}\gtrsim 1470$, a single slope 
is observed in the lifetime statistics, which indicates that B$_1$ 
and B$_2$ have merged to form B. We further reduced the Reynolds 
number in an attempt to detect the initiation of the merger, but this 
resulted in the sudden disappearance of B at around $\textrm{Re}\sim 
1450$, suggesting that, unlike what happens with A and B at higher 
$\textrm{Re}$, the dynamics associated with the independent B$_1$ and 
B$_2$ saddles is highly unstable and only becomes detectable upon 
their merging to form B. Altogether, the dynamics in the range 
$1460<\textrm{Re}<1540$ is extremely complex and beyond the scope of 
this work. 

\begin{figure} \centering 
\begin{tabular}{c} 
\includegraphics[width=0.72\linewidth,clip=]{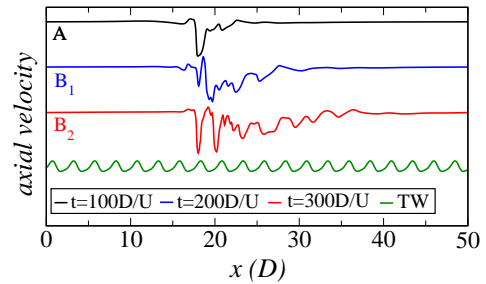} 
\end{tabular} 
\caption{Axial velocity distribution along a 
streamwise line at $(r,\theta)=(0.2\,D,0)$ of the flow snapshots in 
Fig.~\ref{fig:complexity}c and the traveling wave $N2$ found by 
Pringle~\etal~\cite{Pringle2009} ($\textrm{Re}=2400$, axial 
wavelength $2.513\,D$)} \label{fig:loc} 
\end{figure} 

Recently, Chantry \emph{et al}.~\cite{chantry2014} showed that the 
localised relative periodic orbit underlying saddle A emerges from a 
spatially periodic travelling wave (named $N2$) at a subharmonic Hopf 
bifurcation. Figure~\ref{fig:loc} shows the streamwise velocity along 
an axial line for $N2$ and for the flow snapshots of 
Fig.~\ref{fig:complexity}c. The internal wave structure of snapshots 
of B$_1$ and B$_2$, consisting of wave trains of different lengths, 
is also strikingly reminiscent of $N2$. Despite this similarity we 
could not determine whether exact solutions underlying B$_1$ and 
B$_2$ also stem from $N2$ or other traveling-wave solutions. A 
Newton-Krylov solver specially devised to compute relative periodic 
solutions invariably failed to converge from any of the many 
snapshots of B$_1$ and B$_2$ that we seeded as initial guesses. 

In planar shear flow spanwise localisation has been shown to follow 
homoclinic snaking~\cite{burke2007}, which generates multiple 
spanwise-localised solutions spanning different counts of 
vortex-pairs~\cite{schneider2010}. Although an analogous mechanism 
may generate localised relative periodic orbits that underlie A, B$_1$ 
and B$_2$, \citet{mellibovsky2015} have recently demonstrated in the 
case of channel flow that the snaking mechanism is decisively 
disrupted in the streamwise direction. It was found instead that 
streamwise-localised relative periodic orbits of quantised lengths, 
exactly as the chaotic spots found here, appear in saddle-node 
bifurcations and extend to very high $\textrm{Re}$. These solutions 
are related by intricate connections in parameter space and by 
heteroclinic connections in phase space. We speculate that this type 
of solutions might set the stage for bifurcation cascades into 
chaotic attractors, and for the basin boundary metamorphoses and 
crises that finally result in the complexity of wall-bounded 
turbulent flows. 
 
\section{Conclusion} 

Several routes to chaos were proposed in the seventies to account for 
the emergence of disordered dynamics in systems far from 
equilibrium~\cite{ruelle1971,*newhouse1978,feigenbaum1978,*pomeau1980,
cross1993}. These routes were first illustrated in simple dynamical 
systems and later observed in fluid experiments featuring linearly 
unstable laminar flows~\cite{gollub1975,libchaber1982,*eckmann1981}. 
More recently, numerical studies in small computational domains have 
shown that the same routes apply to linearly stable 
flows~\cite{kreilos2012,*shimizu2014,zammert2015,altmeyer2015}, 
albeit with decisive differences. First, the bifurcation cascade 
leading to chaos cannot start from laminar flow, and stems instead 
from disconnected invariant solutions originated at saddle-node 
bifurcations. These solutions are typically unstable in full space, 
so that all aforementioned numerical investigations rely on 
restricting the dynamics to symmetry subspaces that stabilise them. A 
second essential difference is that shear flow turbulence is 
transient, which requires a transition from permanent to transient 
chaos as $\textrm{Re}$ increases. While many numerical studies of 
shear flows in small domains have demonstrated the emergence of 
chaotic saddles, the associated transients invariably produce 
decreasing lifetimes for increasing $\textrm{Re}$. Experimental 
turbulent lifetimes rapidly surge instead, such that additional 
qualitative changes must take place in phase space. Finally, 
transitional shear flow turbulence is spatially localised and it is 
only through spatial proliferation of spots that turbulence becomes 
self-sustained~\cite{avila2011,shi2013}. None of the theoretical and 
numerical studies cited takes this important spatial aspect into 
consideration, the former using models based on low-dimensional ODEs 
and maps, the latter employing numerical domains too small to capture 
localisation. 

\begin{figure} \centering \begin{tabular}{c} (a)\\ 
\includegraphics[width=0.84\linewidth,clip=]{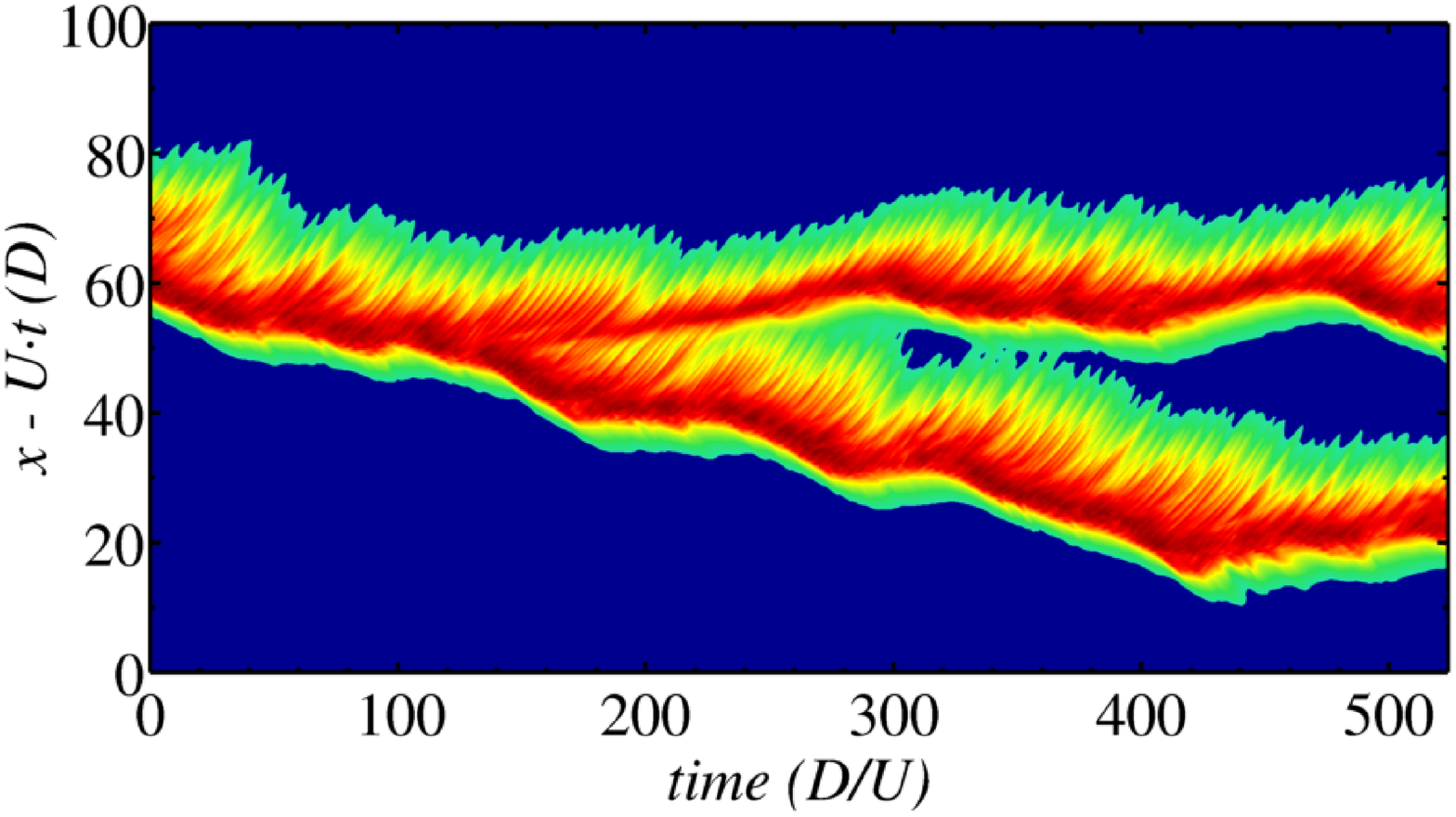}\\ 
(b)\vspace{1mm}\\ \begin{tabular}{c} $t=150\,D/U$\\ 
\includegraphics[width=\linewidth,clip=]{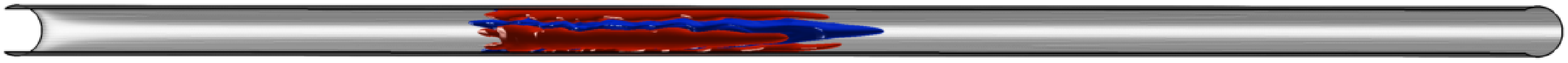}\\ 
$t=400\,D/U$\\ 
\includegraphics[width=\linewidth,clip=]{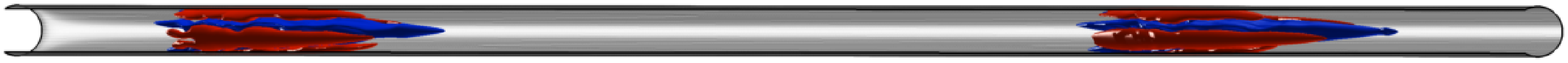}\\ 
\end{tabular} \end{tabular} \caption{(a) Space-time diagram 
and (b) streamwise velocity isosurfaces at $\textrm{Re}=1800$ showing 
a splitting event; Axes and colouring as in Fig.~\ref{fig:puff}.} 
\label{fig:split} \end{figure}

The first reported bifurcation cascade in a spatially extended shear 
flow, Hagen-Poiseuille flow in a long cylindrical 
pipe~\cite{avila2013}, laid out a two-step transition of a streamwise 
localised relative periodic orbit into a localised chaotic spot. Here 
we have shown that a boundary crisis in which the attractor collides 
with its own basin boundary is responsible for the transition from 
permanent to transient chaos. After the boundary crisis, lifetimes 
decrease rapidly with increasing $\textrm{Re}$ following the simplest 
possible scaling law in low dimensional dynamical 
systems~\cite{grebogi1986}. The dynamics of the resulting transient 
spot is only mildly chaotic, involving but weak fluctuations of 
pressure drop, propagation speed and length, typical of merely 
temporal chaos. We have shown that strong spatio-temporal chaotic 
fluctuations arise from a progressive merger of temporally chaotic 
saddles in phase space. Each saddle is associated with a chaotic spot 
of distinct characteristic streamwise length, propagation speed and 
pressure drop, such that their merger results in a spot whose length 
oscillates erratically. Completion of the saddles merger is followed 
at higher $\textrm{Re}$ by phenomena typically observed in 
experiments~\cite{hof2006} such as rapid lifetime growth with 
increasing Reynolds number (see Fig.~\ref{fig:merger}b) or the 
enhancement of spatio-temporal fluctuations through an increase in 
the transition rate between the phase space regions originally 
occupied by isolated saddles. Overall, the dynamics after the merger 
is consistent with that of the turbulent puff. Furthermore, a spatial 
mechanism of occasional violent growth appears to be a precursor to 
splitting, which we have observed to start occurring at larger 
$\textrm{Re}\gtrsim 1700$ (see Fig.~\ref{fig:split}). Increasing 
lifetimes and spatial expansion of spots set the basis for an analogy 
between turbulent transition and directed 
percolation~\cite{pomeau1986,lemoult2016}, thus rendering our work 
relevant to future development of a rational foundation for this 
phenomenology. 

A multitude of streamwise periodic travelling waves with different 
symmetries have been found in the last decade. These waves only 
require subharmonic bifurcations that have been shown to be 
generic~\cite{chantry2014} to produce streamwise localised solution 
branches that could, in turn, give rise to temporally chaotic 
saddles. Hence, analogous scenarios will most probably be at play 
within symmetry subspaces other than that considered here. Piecing 
together the progress made in different subspaces in order to explain 
the complexity observed in full space remains however an outstanding 
challenge. It is unclear as of today how to tackle the problem in the 
abscence of symmetry restrictions given that the localised edge state 
in full space is chaotic at all Reynolds numbers hitherto 
investigated \cite{mellibovsky2009,duguet2010}. A quantitative 
characterisation of the progressive complexification of the dynamics, 
namely through the analysis of the evolution of the Lyapunov spectrum 
of spatio-temporal transients (see \cite{kaneko1989} for an 
application to coupled-map lattices), is beyond the scope of this 
work and therefore left for future study. 

Beyond the example of pipe flow, we expect our findings to be 
relevant also for flows with two spatially extended directions, such 
as channel and Couette flows. In these cases turbulence appears in 
the form of patches that are localised in both the streamwise and 
spanwise directions~\cite{emmons1951} and that feature yet richer 
spatial self-organisation processes~\cite{duguet2013}. Despite the 
challenge, doubly-localised invariant solutions have recently been 
discovered in these geometries~\cite{brand2014,*zammert2014} and are 
promising candidates for building blocks of turbulent dynamics in 
doubly-extended systems. More generally, the mechanisms shown here 
may be relevant to other extended systems driven far from 
equilibrium. In particular, we expect our route to complexity to be 
generic in excitable media, for which theories of temporal chaos also 
fail to properly characterise the spatio-temporal dynamics observed.

\begin{acknowledgments}Support from the Deutsche Forschungsgemeinschaft (DFG) through grant FOR 1182 and computing time from the Jülich Supercomputing Centre (grant number HER22) and Regionales Rechenzentrum Erlangen (RRZE) are acknowledged. 
FM (Serra H\'unter fellow) also acknowledges financial support from 
grants FIS2013-40880-P and 2014SGR1515 from the Spanish and Catalan 
governments, respectively. \end{acknowledgments}

\end{document}